\documentclass[12pt, fleqn]{article}
\usepackage[mathcal]{euscript}

\usepackage{graphicx}
\usepackage{amssymb}
\usepackage{amsmath}
\usepackage{bm} 
\usepackage{epstopdf}
\usepackage{latexsym}
\usepackage[pdftex,bookmarks,breaklinks]{hyperref}
\hypersetup{colorlinks,%
	citecolor=blue,
	}
\usepackage[all]{xy}
\usepackage[pdftex]{}

\usepackage{indentfirst}

\usepackage{marvosym}

\def\K3{\mathrm K3}

\def\double #1{#1{\hbox{\kern-2pt $#1$}}}

\def\pp{{\mathchoice
         {
             \kern 1pt%
             \raise 1pt
             \vbox{\hrule width5pt height0.4pt depth0pt
                   \kern -2pt
                   \hbox{\kern 2.3pt
                         \vrule width0.4pt height6pt depth0pt
                         }
                   \kern -2pt
                   \hrule width5pt height0.4pt depth0pt}%
                   \kern 1pt
          }
           {
             \kern 1pt%
             \raise 1pt
             \vbox{\hrule width4.3pt height0.4pt depth0pt
                   \kern -1.8pt
                   \hbox{\kern 1.95pt
                         \vrule width0.4pt height5.4pt depth0pt
                         }
                   \kern -1.8pt
                   \hrule width4.3pt height0.4pt depth0pt}%
                   \kern 1pt
           }
           {
             \kern 0.5pt%
             \raise 1pt
             \vbox{\hrule width4.0pt height0.3pt depth0pt
                   \kern -1.9pt  
                   \hbox{\kern 1.85pt
                         \vrule width0.3pt height5.7pt depth0pt
                         }
                   \kern -1.9pt
                   \hrule width4.0pt height0.3pt depth0pt}%
                   \kern 0.5pt
           }
           {
             \kern 0.5pt%
             \raise 1pt
             \vbox{\hrule width3.6pt height0.3pt depth0pt
                   \kern -1.5pt
                   \hbox{\kern 1.65pt
                         \vrule width0.3pt height4.5pt depth0pt
                         }
                   \kern -1.5pt
                   \hrule width3.6pt height0.3pt depth0pt}%
                   \kern 0.5pt
           }
       }}

\def\mm{{\mathchoice
                      {
                            \kern 1pt
              \raise 1pt    \vbox{\hrule width5pt height0.4pt depth0pt
                                 \kern 2pt
                                 \hrule width5pt height0.4pt depth0pt}
                            \kern 1pt}
                      {
                           \kern 1pt
              \raise 1pt \vbox{\hrule width4.3pt height0.4pt depth0pt
                                 \kern 1.8pt
                                 \hrule width4.3pt height0.4pt depth0pt}
                            \kern 1pt}
                      {
                           \kern 0.5pt
              \raise 1pt
                           \vbox{\hrule width4.0pt height0.3pt depth0pt
                                 \kern 1.9pt
                                 \hrule width4.0pt height0.3pt depth0pt}
                           \kern 1pt}
                      {
                          \kern 0.5pt
            \raise 1pt  \vbox{\hrule width3.6pt height0.3pt depth0pt
                                 \kern 1.5pt
                                 \hrule width3.6pt height0.3pt depth0pt}
                          \kern 0.5pt}
                      }}

\def\ad{{\kern0.5pt
                  \alpha \kern-5.05pt
\raise5.8pt\hbox{$\textstyle.$}\kern 0.5pt}}

\def\bd{{\kern0.5pt
                  \beta \kern-5.05pt \raise5.8pt\hbox{$\textstyle.$}\kern 0.5pt}}

\def\qd{{\kern0.5pt
                  q \kern-5.05pt \raise5.8pt\hbox{$\textstyle.$}\kern 0.5pt}}
\def\Dot#1{{\kern0.5pt
    {#1} \kern-5.05pt \raise5.8pt\hbox{$\textstyle.$}\kern 0.5pt}}

\def\a{\alpha}
\def\ad{\dot\a}
\def\b{\beta}
\def\bd{\dot\b}
\def\g{\gamma}

\def\d{\delta}

\def\s{\sigma}

\def\t{\theta}
\def\l{\lambda}
\def\o{\omega}
\def\p{\partial}
\def\pb{\overline\p}
\def\be{\begin{eqnarray}}
\def\ee{\end{eqnarray}}

\def\e{{\epsilon}}
\def\ve{{\varepsilon}}
\def\N{{\nabla}}
\def\st{{\widetilde\s}}

\def\ket#1{\left| #1\right>}

\begin{document}
~
\vspace{-110pt}

\begin{flushright}
\makebox[0pt][b]{}
\end{flushright}

\vspace{40pt}

\center{{\LARGE Compactification of the Heterotic Pure Spinor\\ Superstring II}

\vspace{20pt}

{\large Osvaldo~Chand\'ia$,\hspace{-9pt}{}^{\small \mbox\Aries}$\
William~D.~Linch~{\sc iii}$,\hspace{-6pt}{}^{\small \mbox\Pisces}$\
and Brenno Carlini Vallilo$.\hspace{-4pt}{}^{\small \mbox \Libra}$ }
}

\vspace{-10pt}

\center{
${}^{\small \mbox\Aries}${\em
Departamento de Ciencias, Facultad de Artes Liberales\\
\& Facultad de Ingenieria y Ciencias, \\
Universidad Adolfo Iba\~nez,\\
Santiago de Chile.}

\vspace{-10pt}

${}^{\small \mbox\Pisces,\mbox\Libra}${\em
Departamento de Ciencias F\'isicas,\\
Facultad de Ciencias Exactas,\\
Universidad Andres Bello,\\
Santiago de Chile.}
}

\abstract{We study compactifications of the heterotic pure spinor superstring to six and four dimensions focusing on two simple Calabi-Yau orbifolds. We show that the correct spectrum can be reproduced only if, in the twisted sector, there remain exactly 5 and 2 pure spinor components untwisted, respectively. This naturally defines a ``small'' Hilbert space of untwisted variables. We point out that the cohomology of the reduced differential on this small Hilbert space can be used to describe the states in the untwisted sector, provided certain auxiliary constraints are defined. In dimension six, the mismatch between the number of pure spinor components in the small Hilbert space and the number of components of a six-dimensional pure spinor is interpreted as providing the projective measure on the analytic subspace (in the projective description) of harmonic superspace.}

\begin{flushleft}
~\\
{${}^{\small \mbox\Aries}$\href{mailto:ochandiaq@gmail.com}{ochandiaq@gmail.com}}\\
{${}^{\small \mbox\Pisces}$ \href{mailto:wdlinch3@gmail.com}{wdlinch3@gmail.com}}\\
{${}^{\small \mbox\Libra}$ \href{mailto:vallilo@gmail.com}{vallilo@gmail.com}}
\end{flushleft}

\setcounter{page}0
\thispagestyle{empty}

\newpage
\tableofcontents

\section{Introduction}

The pure spinor formalism is well-developed in ten flat space-time dimensions. Although its quantization procedure is not fully understood, it has passed many consistency tests.
Its applications to curved backgrounds, however|$AdS_5\times S^5$ being an exception|remains much less explored. In a previous paper \cite{Chandia:2009it}, we discussed the compactification
of the heterotic string in a Calabi-Yau 3-fold background. We showed how cohomology of the Berkovits differential and supersymmetry conditions fix the internal geometry in the expected way. Although straightforward, there had previously been various unsuccessful attempts to reproduce this result using some reasonable replacement of the ten-dimensional pure spinor \cite{group}. In these attempts, the off-shell spectrum of superfields is reproduced with ease but the auxiliary field equations of motion are missing. While the analysis using all components of the dimensionally reduced pure spinor resolves this issue, the disadvantage of this solution is that the spectrum does not emerge in terms of irreducible superfield representations of the lower-dimensional space-time symmetry group. In this sense, it is not a fully satisfactory compactification.

In this paper, we attempt to shed light on this compactification problem by studying simple orbifold compactifications to six and four dimensions preserving eight and four supercharges. The familiar splitting of the resulting spectrum into twisted and untwisted sectors is particularly helpful in guiding the analysis. Specifically, we will show that the known spectra cannot be reproduced unless the number of untwisted pure spinors in the twisted sector is $n=5$ and $n=2$, respectively. This should be contrasted with the dimensions of the pure spinor representations in $D=6$ and $D=4$ which are $4$ and $2$, respectively \cite{Berkovits:2005hy}.

The picture that emerges is that the spectrum may be computed in a reduced cohomology on a ``small'' Hilbert space consisting of only the untwisted variables. The reduced differential serves to define the superfield representation which houses these components. For the fields coming from the untwisted sector, the translation from components to superfields requires the introduction of additional constraints. This is possible due to the reduced symmetry of the compactified string and necessary since the off-shell superfield formalism naturally introduces auxiliary components.

In this work, we will content ourselves with the insights gained from mostly space-time arguments. The full understanding of the results requires a deeper study of the conformal field theory of the orbifolded pure spinor which is beyond the scope of this paper.

\section{Ten dimensions}
In this section we review supersymmetry in ten dimensions and the pure spinor formalism in this case \cite{Berkovits:2000fe}. Our ten-dimensional conventions are summarized in appendix \ref{appendix10}. Consider the supersymmetry algebra with sixteen supercharges in ten dimensions
\begin{eqnarray}
\label{susy10d}
{ \{ } D_\alpha , D_\beta { \} } = -2 i \gamma^m_{\alpha \beta} \partial_m~,
\end{eqnarray}
where $m=0, \dots, 9, \a, \b=1, \dots, 16$ and $\g^m$ are the $16\times 16$ symmetric gamma matrices.

The super-Maxwell system is described by a Grassmann superfield $A_\a(X,\t)$ constrained by $\g_{mnpqr}^{\a\b}D_\a A_\b=0$. In order to obtain the equations of motion of the ten-dimensional super-Maxwell fields, we define potential superfields from the above constrained one. We define a vector superfield $A_m$ as
\begin{eqnarray}
\label{am10}
D_{(\a} A_{\b)} = -2i \g^m_{\a\b} A_m~.
\end{eqnarray}
Similarly, a Grassmann superfield $W^\a$ is defined according to
\begin{eqnarray}
\label{dam}
D_\a A_m-\p_m A_\a = -2i(\g_m)_{\a\b} W^\b~.
\end{eqnarray}
The superfields $A_m$ and $W^\a$ turn out to be related as
\begin{eqnarray}
\label{dW}
D_\a W^\b=\frac{1}{4}(\g^{mn})_\a{}^\b F_{mn}~,
\end{eqnarray}
where $F_{mn}=\p_{[m}A_{n]}$. All these relations imply
\begin{eqnarray}
\label{maxwell}
\p^m F_{mn}=0,\quad (\g^m)_{\a\b} \p_m W^\b=0~,
\end{eqnarray}
which state that the photon is the lowest $\t$-component of $A_m$ and the photino is the lowest $\t$-component of $W^\a$.

The pure spinor string describes the above system in a manifestly supersymmetric way. The idea, originally due to Siegel \cite{Siegel:1985xj}, is to use the superspace coordinates as free variables in a world-sheet action given by
\begin{eqnarray}
\label{s10}
S=\int d^2z~\frac{1}{2}\p X^m \pb X_m + p_\a \pb\t^\a +\cdots~,
\end{eqnarray}
where $p_\a$ is the variable canonically conjugate to $\t^\a$ and the ellipsis denotes the pure spinor contribution to the action, the explicit form of which is not needed.

Physical states are defined in the cohomology of the nilpotent operator
\begin{eqnarray}
\label{qps}
Q=\oint \l^\a d_\a~,
\end{eqnarray}
where $\l^\a$ is the pure spinor variable constrained by
\be
\l\g^m\l=0~,
\ee
and $d_\a$ is the world-sheet representation of the superspace derivative $D_\a$ given by
\begin{eqnarray}
\label{da}
d_\a=p_\a+i\g^m_{\a\b}\t^\b\p X_m + \frac{1}{2}\g^m_{\a\b}(\g_m)_{\g\d} \t^\b \t^\g \p\t^\d~.
\end{eqnarray}
Note that $Q$ is nilpotent because of the pure spinor condition and the operator product
\begin{eqnarray}
\label{dd}
d_\a(y) d_\b(z) \to \frac{2i}{(y-z)} \g^m_{\a\b} \Pi_m(z)~,
\end{eqnarray}
where
\be
\label{pim}
\Pi^m=\p X^m-i(\t\g^m\p\t)~.
\ee

The unintegrated massless vertex operator is given by $U=\l^\a A_\a$. The condition $QU=0$ puts the superfield $A_\a$ on-shell. It is interesting to note that the integrated vertex operator, necessary to compute scattering amplitudes, is defined to satisfy $QV=\p U$. That is,
\be
\label{vint}
V=\p\t^\a A_\a+\Pi^m A_m+d_\a W^\a+\frac{1}{2}N^{mn}F_{mn}~,
\ee
where $N^{mn}=\frac{1}{2}(\l\g^{mn}\o)$ with $\o$ being the canonical conjugate variable of the pure spinor field.

\paragraph{Spectrum}
The 0-momentum cohomology of the ten-dimensional string is concentrated in levels $\lambda^p \theta^q$ with $(p,q)=(0,0), (1,1), (1,2),$ $ (2, 3),(2,4), (3, 5)$.
The $p=1$ cohomology is generated by
\begin{eqnarray}
\Gamma^m = \lambda \gamma^m \theta
	~~~\mathrm {and}~~~
\Gamma_\alpha = \Gamma^m (\gamma_m \theta)_\alpha~,
\end{eqnarray}
corresponding to the field and ino. The cohomology with $p=2$ is generated by
\begin{eqnarray}
\Gamma^*_m = \Gamma^\alpha \gamma^m_{\alpha \beta}\Gamma^\beta
	~~~\mathrm {and}~~~
\Gamma^{*\alpha} =\Gamma_m (\gamma^{m\alpha \beta} \Gamma_\beta)~,
\end{eqnarray}
corresponding to the anti-field and anti-ino. The top level of the cohomology ($p=3$) is generated by
\begin{eqnarray}
\label{topform}
\Gamma_\alpha \Gamma^{*\alpha} =
\Gamma^*_m (\Gamma_\alpha \gamma^{m\alpha \beta} \Gamma_\beta)=
\Gamma_m\Gamma^{*m}~.
\end{eqnarray}

\section{Six dimensions}
In this section we study the six-dimensional compactification of the pure spinor string.
There are two ways to proceed. The most straightforward is the approach taken previously for compactifications to four dimensions \cite{Chandia:2009it}.
As in ten dimensions, one writes the pure-spinor-number 1 massless unintegrated vertex operator
\begin{eqnarray}
U= \lambda^{\alpha a} A_{\alpha a}+ \lambda_{\alpha a^\prime} A^{\alpha a^\prime}~
\end{eqnarray}
and computes the cohomology of the Berkovits differential
\begin{eqnarray}
\label{Q}
Q= \oint\left( \lambda^{\alpha a} d_{\alpha a}  + \lambda_{\alpha a^\prime}d^{\alpha a^\prime}\right)~,
\end{eqnarray}
giving the dimensionally reduced pure spinor constraints on the potentials $A$. Checking Bianchi identities up to dimension 2 shows that the condition to have $N=(1,0)$ target space supersymmetry is that the compactification manifold should have a curvature form of definite duality and that the gauge bundles should have holomorphic connections with the same duality. These are the six-dimensional analogues of the $F$- and $D$-term conditions in four-dimensions.

Alternatively, we perform an orbifold operation directly on the ten-dimensional pure spinor string.
A complication arises in the twisted sector due to the constraint on the pure spinor. Our knowledge of the spectrum uniquely determines the left-moving part of the twisted vacua.

\subsection{Supersymmetry in six dimensions}

Consider compactification to six dimensions.\footnote{Our six-dimensional conventions and various relevant identities are summarized in part \ref{6dsusy} of the appendix.} The bosonic superspace coordinates split as $(X^m, X^i)$ where $m=0, \dots, 5$ and $i=6, \dots, 9$. The fermionic superspace coordinates are $(\t^{\a a}, \t^{a'}_\a)$ where $\a=1, \dots, 4$ are $SU(4)$ spinor indices and $a, a'=1,2$ are $SU(2)\times SU(2)$ spinor indices \cite{Koller:1982cs}.
The pure spinor variables $(\l^{\a a}, \l^\a_{a'})$ are constrained by
\be
\label{ps6}
\l^{\a a} \l^\b_a+\frac{1}{2}\ve^{\a\b\g\d}\l^{a'}_\g \l_{\d a'}=0,\quad \l^{\a a} \l_\a^{b'}=0~.
\ee

As in ten dimensions, the pure spinor condition allows to define the nilpotent charge (\ref{qps}) which turns out to be the BRST charge of the superstring. Acting on massless states, the BRST charge determines the super-Maxwell equations of motion. In our case, the vector superfields $(A_m, A_i)$ are related to the fermionic potential superfields $(A_{\a a}, A^\a_{a'})$ as
\be
\label{daAb}
D_{\a a}A_{\b b}+D_{\b b}A_{\a a}=-2i\e_{ab}(\s^m)_{\a\b}A_m ~,\\
\label{dapAbp}
D^\a_{a'}A^\b_{b'}+D^\b_{b'}A^\a_{a'}=-2i\e_{a'b'}(\s^m)^{\a\b}A_m~,\\
\label{daAbp}
D_{\a a}A^\b_{b'}+D^\b_{b'}A_{\a a}=-2i\d_\a^\b(\s^i)_{ab'}A_i~,
\ee
where
\be
\label{da6}
D_{\a a}=\frac{\p}{\p\t^{\a a}}-i(\s^m)_{\a\b}\t^\b_a\p_m-i(\s^i)_{ab'}\t^{b'}_\a\p_i~,\cr
\label{dap6}
D^\a_{a'}=\frac{\p}{\p\t^{a'}_\a}-i(\s^m)^{\a\b}\t_{\b a'}\p_m-i(\s^i)_{ba'}\t^{\a b}\p_i~,
\ee
are the superspace covariant derivatives.

The equations (\ref{daAb} -- \ref{daAbp}) imply the existence of a Grassmann superfield whose $\theta$-independent part is the photino of the super-Maxwell multiplet. In order to obtain this superfield, we define the fermionic covariant derivatives $\N_{\a a}=\p_{\a a}+A_{\a a}, \N^\a_{a'}=\p^\a_{a'}+A^\a_{a'}$ and note that they are constrained to satisfy
\be
\label{DaDa}
\{ \N_{\a a}, \N_{\b b} \} = -2i\e_{ab}(\s^m)_{\a\b}\N_m~,\cr
\label{DapDap}
\{ \N^\a_{a'},\N^\b_{b'} \}=-2i\e_{a'b'}(\s^m)^{\a\b}\N_m~,\cr
\label{DaDap}
\{ \N_{\a a},\N^\b_{b'} \}=-2i\d_\a^\b(\s^i)_{ab'}\N_i~,
\ee
where $\N_m=\p_m+A_m, \N_i=\p_i+A_i$. The Bianchi identities involving these fermionic covariant derivatives imply the existence of the dimension-$\frac32$ field strengths
\be
\label{DaD}
[\N_{\a a},\N_m]=-2i(\s_m)_{\a\b}W^\b_a ~,\quad [\N_{\a a},\N_i]=-2i(\s_i)_{ab'}W^{b'}_\a~,\\
\label{DapD}
[\N^\a_{a'},\N_m]=-2i(\s_m)^{\a\b} W_{\b a'}~, \quad [\N^\a_{a'},\N_i]=-2i(\s_i)_{ba'}W^{\a b}~.
\ee
Note that there are two ways to write each field strength. This fact allows the derivation of relations between {\it a priori} independent superfield strengths as we now show.

We begin by defining the dimension-2 field strengths
\begin{eqnarray}
{[} \nabla_m , \nabla_n{]} =	 F_{mn}
~,~
{[ }\nabla_{m}, \nabla_{i}{]}= F_{mi}
~,~{[ }\nabla_{i}, \nabla_{j}{]}= F_{ij}~.
\end{eqnarray}
Additionally, there are the superfield strengths
\begin{eqnarray}
d=\{ \nabla_{\alpha a} , W^{\alpha a}\}
	~,~ f_{ab}=\{ \nabla_{\alpha (a} , W^{\alpha}_{ b)}\}~,~
d^\prime = \{ \nabla^{\alpha a^\prime}  ,  W_{\alpha a^\prime}\}
	~,~ f_{a^\prime  b^\prime}= \{ \nabla^\alpha_{ ( a^\prime}  ,W_{\alpha  b^\prime)}\}~.
\end{eqnarray}
However, the dimension-2 Bianchi identities
\begin{eqnarray}
\{ \nabla_{\alpha a} , W^{\beta b}\} &=&
	\delta_a^b (\sigma^{mn})_\alpha{}^\beta F_{mn}+\delta_\alpha^\beta (\sigma^{ij})_a{}^bF_{ij}~,~\cr
\{ \nabla^{\alpha a^\prime}  ,  W_{\beta b^\prime}\} &=&
	\delta_{b^\prime}^{a^\prime}(\sigma^{mn})^\beta{}_\alpha F_{mn}+ \delta_\beta^\alpha (\sigma^{ij})_{b^\prime}{}^{a^\prime}F_{ij}~,~\cr
\{ \nabla_{\gamma a} , W_{\delta a^\prime}\}&=&
	- (\sigma^m)_{\alpha \beta} (\sigma^i)_{a a^\prime}F_{mi}
	~,~
\end{eqnarray}
which follow from equation (\ref{DaD}), can be re-written using equation (\ref{DapD}) instead. This implies that
\begin{eqnarray}
d=0 ~,~ f_{ab} = \frac14 (\sigma^{ij})_{ab} F_{ij} ~,~d^\prime =0 ~,~ f_{a^\prime b^\prime} = \frac14 (\sigma^{ij})_{a^\prime b^\prime} F_{ij}~.
\end{eqnarray}

Consider now a background that preserves $N=1$ supersymmetry in six dimensions. The fermionic superfield, as any superfield, can be expanded in $\t$. In our case, assuming Lorentz invariance of the six-dimensional background, we have
\be
W^{\a a} = \t^{\a b} f_b{}^a + \cdots~, \quad W^{a'}_\a = \t^{b'}_\a f_{b'}{}^{a'} + \cdots~.
\label{Wback}
\ee
Note that only combinations that are even under the orbifold operation, which is introduced in the next subsection, are allowed  in this expansion. In this way, the terms in (\ref{Wback}) are valid. Since the background preserves six-dimensional supersymmetry, the background fields are invariant under shifts of $\t^{\a b}$. Thus, the background field $f_a{}^b$, which is the self-dual part of the vector field strength $F_{ij}$, vanishes. This is the six-dimensional relation between spacetime supersymmetry and the internal geometry. An analysis for the full superstring is entirely analogous to that performed in the four-dimensional case in reference \cite{Chandia:2009it} and implies the usual conditions of Ricci-flatness of the metric and the Hermitian-Yang-Mills equations for the heterotic gauge field.

\subsection{K3 Orbifold}
\label{K3O}
We proceed to reduce the supersymmetry $N=(1,1)\to (1,0)$. Geometrically, this can be achieved by orbifold projection. To begin, we form quaternionic combinations of the internal coordinates
\begin{eqnarray}
\label{H}
\left( X^{aa^\prime}\right) = \left(
	\begin{array}{cc}
	X^6+iX^7 & X^8-iX^9\\
	X^8+iX^9 & -X^6+iX^7
	\end{array}
	\right)~.
\end{eqnarray}
Next, we define the holomorphic coordinates
\begin{eqnarray}
(Z^i)=(X^{i1^\prime})~,~(\bar Z_i)=(\epsilon_{ij}X^{j2^\prime})~,
\end{eqnarray}
where we are identifying the holomorphic $U(2)$ index of $\mathbb C^2$ with the $SU(2)_L \subset Spin(4)$ Lorentz index.

The K3 orbifold we will consider is obtained by quotienting the square 2-torus $T^2= \mathbb C^2/\mathbb Z^2$ by the relation $Z^i\mapsto -Z^i$.
This is implemented by constructing a central element $h=\mathrm e^{i\pi \oint L}\in \mathbb Z_2\subset SU(2)_L\subset Spin(4)$ of the internal Lorentz group where $L=\left(L_{67}\pm L_{89}\right)$ denotes the orbital part of the spin generator.
We will select the negative sign so that the generator of the symmetry $L=L_{67}-L_{89}$ is anti-self-dual and, therefore, of the type $L^{a^\prime b^\prime}= -\epsilon_{ab} \, X^{aa^\prime} \! {\stackrel{\leftrightarrow}\partial}\! \,X^{bb^\prime} = L^{b^\prime a^\prime}$. Checking commutators such as
${ [ }{\oint} L , Z^1 { ] } = { [ } L_{67} , X^6+iX^7 { ] } = i X_{[6}\partial_{7]} (X^6+i X^7)
= - Z^1$
and
${ [ } {\oint}L^{1^\prime 2^\prime }, Z^i{ ] } =-  { [ } Z^{j} \! {\stackrel{\leftrightarrow}\partial}\! \, \bar Z_{j}, Z^i{ ] } = - Z^i$
shows that, in fact, $L= L^{1^\prime 2^\prime}$. 

The extension to the rest of superspace is fixed by the differential: Since ${ [ } Q, L^{1^\prime 2^\prime}{ ] }\neq 0$, additional Grassmann ($S$) and pure spinor terms ($N$) are needed to make $M^{1^\prime 2^\prime}= L^{1^\prime 2^\prime}+S^{1^\prime 2^\prime}+N^{1^\prime 2^\prime}$ in the cohomology of the Berkovits differential.
Explicitly,
\begin{eqnarray}
S^{ab}=\t^{\a (a}p_\a^{b)}~,~ S^{a'b'}=\t_\a^{(a'}p^{\a b')}~,~
N^{ab}=\l^{\a (a}\o_\a^{b)}~,~ N^{a'b'}=\l_\a^{(a^\prime}\o^{\a b')}~.
\end{eqnarray}
This implies that ${ [ } \oint M^{1^\prime 2^\prime}, \theta^{\alpha a} { ] } = (-)^{a} \theta^{\alpha a}$ and ${ [ } \oint M^{1^\prime 2^\prime}, \theta_\alpha^{a'} { ] } = 0$ and similarly for the pure spinor.
It follows that the central element
\begin{eqnarray}
\mathcal M =\mathrm e^{ i \pi \oint M^{1^\prime 2^\prime}}~,
\end{eqnarray}
equals $-1$ on all combinations of worldsheet variables with an uncontracted $a^\prime$-index with the result that the untwisted sector lives in $N=(1,0)$ superspace.

With the choice (\ref{H}) of isomorphism $\mathbb R^4 \cong \mathbb C^2$ we are selecting a point in the coset $SO(4)/U(2)\cong S^2$ of orthogonal complex structures on $\mathbb R^4$. Let us parameterize this choice with a harmonic variable $u_{a^\prime}$ \cite{Galperin:2001uw} and extend to the cohomology of the differential by defining $\Gamma^*:= iQ(X^*)$. This gives
\begin{eqnarray}
\begin{array}{ll}
\Gamma^{\alpha \beta} =
	\lambda^{[\alpha a} \theta^{\beta]}_a +\tfrac12 \epsilon^{\alpha \beta \gamma \delta} \lambda_{\gamma a^\prime} \theta_\delta^{a^\prime}~, &
\Gamma_{\alpha i} = \Gamma_{\alpha \beta} \theta^{\beta}_i + \Gamma_{ia^\prime} \theta_{\alpha}^{ a^\prime}~,\\
\Gamma^{i} = - \lambda^{\alpha i} \theta_{\alpha} - \lambda_{\alpha} \theta^{\alpha i}~, &
\Gamma^{\alpha} = \Gamma^{i} \theta^\alpha_i - \Gamma^{\alpha \beta} \theta_\beta~.
\end{array}
\end{eqnarray}
where
\begin{eqnarray}
\lambda_\alpha = u^{a^\prime} \lambda_{\alpha a^\prime}~~\mathrm{ and }~~\theta_\alpha = u^{a^\prime} \theta_{\alpha a^\prime}~.
\end{eqnarray}
This superspace was used in \cite{Sokatchev:1995nj} to construct an action principal for self-dual Yang-Mills.
It is fitting that it should appear in the pure spinor compactification on K3.

To recapitulate, we seek to derive the result of orbifolding the 4-torus by $\mathbb Z_2$ acting on the {\em right-moving} side as
\begin{eqnarray}
\mathcal M(X_R^a,Z_R^i, \rho^{\mathrm i} , \rho^I, \rho^\mathbf I)
	= (X_R^a,- Z_R^i, - \rho^{\mathrm i} , \rho^I, \rho^\mathbf I)~,
\end{eqnarray}
and on the {\em left-moving} side as
\begin{eqnarray}
\mathcal M(X_L^a,Z_L^i,\theta^{\alpha a} , \theta_{\alpha} , \lambda^{\alpha a} , \lambda_{\alpha })
	= (X_L^a, - Z_L^i,\theta^{\alpha a} , - \theta_{\alpha} , \lambda^{\alpha a} ,- \lambda_{\alpha})~,
\end{eqnarray}
where $I=1,\dots, 12$ and $\mathbf I = 1, \dots 16$.

\paragraph{Untwisted sector} The physics of the {\em right-moving} side is identical to the RNS analysis. The embedding of the spin connection into the gauge connection breaks the first $E_8$ factor of the gauge group. The breaking and the resulting decomposition of the adjoint representation are given by
\begin{eqnarray}
E_8& \longrightarrow& SU(2)\times E_7~\cr
\mathbf {248} &\mapsto& (\mathbf 3, \mathbf 1)\oplus (\mathbf 1, \mathbf {133})\oplus (\mathbf 2, \mathbf {56})~.
\end{eqnarray}
The first two factors are even under the $\mathbb Z_2$ action while the last is odd.
With this, the relevant states are built from\footnote{The adjoint representation is constructed from the states
\begin{eqnarray}
\rho^{\mathrm i}_{-1/2} \rho^{\bar {\mathrm \i}}_{-1/2}\ket0~,~
\rho^{I}_{-1/2} \rho^{J}_{-1/2}\ket0~,~
\rho^{\mathbf I}_{-1/2} \rho^{\mathbf J}_{-1/2}\ket0~
\end{eqnarray}
in the adjoint representation $(\mathbf 3,\mathbf 1,\mathbf 1)\oplus(\mathbf 1,\mathbf {66},\mathbf 1)\oplus(\mathbf 1,\mathbf 1,\mathbf {120})$ of $SU(2)\times Spin(12)\times Spin(16)$ together with the relevant spinors. For the $E_8$ factor this is simply the $Spin(16)$ spinor $\mathbf {128}$. For the $E_7$ factor the situation is more complicated. The decomposition of the adjoint under $Spin(12)$ is
\begin{eqnarray}
\mathbf{133} = \mathbf{66}\oplus \mathbf{32_s}\oplus{\mathbf{32_c}}\oplus \mathbf{2}\oplus \mathbf 1~.
\end{eqnarray}
}
\begin{eqnarray}
\begin{array}{ccccl}
+1 &:& \tilde \alpha_{-1}^{\alpha \beta}\ket0 &,& \ket{a}:a \in (\mathbf 3,\mathbf 1,\mathbf 1)\oplus(\mathbf 1,\mathbf {133},\mathbf 1)\oplus(\mathbf 1,\mathbf 1,\mathbf {248})~,\\
-1 &:& \tilde\alpha_{-1}^{i}\ket0,\tilde\alpha_{-1}^{\bar \imath}\ket0 &,& \ket{f}:f \in (\mathbf 2,\mathbf {56},\mathbf 1)~.\\
\end{array}
\end{eqnarray}
On the {\em left-moving} side, the 0-momentum cohomology generators are
\begin{eqnarray}
\begin{array}{cclcl}
+1&;&\Gamma^{\alpha \beta} \ket0
	 &,&\Gamma_{\alpha i}\ket0~, \\
-1&;&
	\Gamma^{i}\ket0
	 &,&\Gamma^{\alpha}\ket0~. \\
\end{array}
\end{eqnarray}

This cohomology is, of course, isomorphic to that obtained directly from the pure spinor superspace in the previous subsection. More importantly, it agrees with a reduced cohomology with differential
\begin{eqnarray}
\label{Q0}
Q_0=\oint \bm\lambda^\alpha d^+_\alpha~,
\end{eqnarray}
in the full harmonic superspace \cite{Galperin:2001uw} provided auxiliary conditions are imposed. Here, $d^+_\alpha$ is the worldsheet current acting by the harmonic derivative $D^+_\alpha$ on harmonic superfields and $\bm \lambda^\alpha$ is an unconstrained six-dimensional spinor. For the vector multiplet in the $\tau$-frame, the harmonic superfield is given by the potential $A^+_\alpha$ in the covariant derivative $\nabla^+_\alpha$. In the $\lambda$-frame, this potential is gauged away but the multiplet re-appears in the form of a connection $V^{++}$ for the harmonic superspace operator $D^{++}$. The auxiliary condition is $F^{++}=0$. For the hypermultiplet, the harmonic superfield is $q^+$ and the auxiliary condition is $D^{++}q^+=0$.

The states surviving the projection are as follows: From the $(+1,+1)$ combinations we obtain
\begin{eqnarray}
\tilde \alpha_{-1}^{\alpha \beta} \Gamma^{\gamma \delta}\ket0 ~,~ \tilde \alpha_{-1}^{\alpha \beta} \Gamma_{\gamma c}\ket0~,
\end{eqnarray}
forming the supergravity multiplet and
\begin{eqnarray}
\Gamma^{\alpha \beta}\ket a ~,~ \Gamma_{\gamma i}\ket a~,
\end{eqnarray}
giving the super-Yang-Mills multiplet in the adjoint of the unbroken gauge group $SU(2)\times E_7\times E_8$. The $(-1,-1)$ combination, again, gives two supermultiplets. The first is
\begin{eqnarray}
\tilde \alpha_{-1}^{\bar \imath} \Gamma^{j} \ket0 ~,~\tilde \alpha_{-1}^{\bar \imath} \Gamma^{\alpha} \ket0~,
\end{eqnarray}
which give the $2\times 2$ $E_7$-neutral scalars and their superpartners. These correspond to what remains of the middle cohomology of $T^4$ on the quotient. The second supermultiplet is
\begin{eqnarray}
\Gamma^{i} \ket f ~,~ \Gamma^{\alpha} \ket f~,
\end{eqnarray}
which give 4 real scalars in the real representation $({\mathbf 2}, \mathbf {56},\mathbf 1)$ together with their superpartners. The multiplet that couples naturally with such representation is a half-hypermultiplet. We can collect them to form two hypermultiplets. The missing hypermultiplets will come from the fixed points.

\paragraph{Twisted sector} There are $2^4=16$ fixed points of the orbifold action. Each of these will carry a twisted-sector vacuum $\ket{0}_v$.
Let us remind ourselves of the 0-point energies of the worldsheet fields as given by the formula \cite{Polchinski:1998rr}
\begin{eqnarray}
\label{ZPE}
E= (-)^F\left[ \frac1{48} -\frac1{16}(2\theta-1)^2\right]~,
\end{eqnarray}
where $F$ is the fermion number of the state which is zero for bosons and one for fermions. For the real periodic boson this is $-\frac1{24}$ and for the real anti-periodic boson $\theta=\frac12$ it is $\frac1{48}$.

On the {\em right-moving} side, we again copy the RNS result.
The formula for the vacuum energy has a spacetime part which is always
\begin{eqnarray}
(6-2)(-\tfrac 1{24})+ 4(\tfrac1{48})=-\tfrac1{12}~.
\end{eqnarray}
Here we are taking into account the contribution of the $bc$ ghosts by counting only 4 of the 6 spacetime oscillators.
There are four sectors coming from the choice of periodic ($P$) and anti-periodic ($A$) boundary conditions for the heterotic fermions parameterizing the root lattice of $E_8\times E_8$. The fermions in the spin connection must have the same periodicity as the fermions in the commutant of $SU(2)\subset E_8$. Thus, there are, {\it a priori}, four sectors.

In the various sectors we calculate the heterotic fermion contributions. The 4 fermions $\rho^{\mathrm i}$ and $\rho^{\bar{\mathrm \i}}$ which were used to embed the spin connection into the gauge connection are twisted while the other $12+16$ heterotic fermions are not. This gives
\begin{eqnarray}
(P,P)&:&-\tfrac1{12}+ 4(-\tfrac 1{48}) +12(\tfrac1{24})+16(\tfrac 1{24})=1~,\cr
(A,P)&:&-\tfrac1{12}+  4(\tfrac 1{24}) +12(-\tfrac1{48})+16(\tfrac 1{24})= \tfrac12~,\cr
(P,A)&:&-\tfrac1{12}+  4(-\tfrac 1{48}) +12(\tfrac1{24})+16(-\tfrac 1{48})=0~,\cr
(A,A)&:&-\tfrac1{12}+  4 (\tfrac1{24}) +12(-\tfrac1{48})+ 16(-\tfrac 1{48})=-\tfrac12~,
\end{eqnarray}
where in the last two lines, we have used that $\rho^{\mathrm i}$ and $\rho^{\bar{\mathrm \i}}$ are both twisted and antiperiodic ({\it i.e.},~periodic overall).

The vacua with positive mass, $(P,P)$ and $(A,P)$, do not contribute to the massless spectrum.
The massless $(P,A)$ vacuum is acted on by $16-4=12$ fermion 0-modes $\rho_0^I$ which give the Pauli matrices of $Spin(12)$ upon quantization. As usual, this implies that the vacuum transforms in an irreducible spinor representation of $Spin(12)$, in this case the (real) ${\mathbf {32}}$. Finally, the $(A,A)$ vacuum contributes the massless states
\begin{eqnarray}
\label{predoublet}
\rho_{-\frac12}^I\ket0_{vR}\sim \mathbf{12}
~,~
\tilde \alpha_{-\frac12}^{\bar \imath} \ket0_{vR}~,
\end{eqnarray}
but these are acted on by $2$ complex fermion 0-modes $\rho^{\mathrm i}_0$ (which are twisted and anti-periodic). These zero modes satisfy an $so(4)\approx su(2)_L \oplus su(2)_R$ gamma matrix algebra. The minimal representations are doublets of $su(2)_L$ or $su(2)_R$. Since the $\rho^{\mathrm i}_0$ zero modes are odd under ${\mathcal M}$, we can choose one of the $su(2)$ factors to be odd and the other to be even. This means that each state in (\ref{predoublet}) is a doublet of the invariant $su(2)$. (This can also be seen by dividing $\rho^{\mathrm i}_0$ into creation and annihilation operators acting on a Clifford vacuum.) Then the first state in (\ref{predoublet}) combines with the ${\mathbf{32}}$ to give the $\mathbf {12}\oplus\mathbf {12}\oplus {\mathbf{32}}=
{\mathbf {56}}$ of $E_7$. Altogether, these left-movers contribute
\begin{eqnarray}\label{twistright}
\ket{(\mathbf 1, {\mathbf {56}},\mathbf 1)}_{vR} + 2\ket{(\mathbf 2, \mathbf 1,\mathbf 1)}_{vR}~,
\end{eqnarray}
for every massless right-moving vacuum state, which we now construct. 

On the {\em left-moving} side, we have $D=6$ untwisted and $10-D=4$ twisted coordinates. There are also $4N=8$ untwisted and $4(4-N)=8$ twisted fermions which come together with their conjugate momenta. Counting in pairs, these contribute $\frac1{12}$ and $-\frac1{36}$. Finally, we will have $n$ untwisted and $11-n$ twisted pure spinors and their momenta giving $-\frac1{12}$ and $\frac1{36}$ in pairs. In all,\footnote{Written in terms of all complex pairs, this becomes
\begin{eqnarray}
E=
{3( -\tfrac1{12})+ 2(\tfrac1{24})}+
{8(\tfrac1{12})+ 8(-\tfrac1{24})}+
{n(-\tfrac1{12})+(11-n)(\tfrac1{24})}
=\left(-\tfrac 1{12}  -\tfrac1{24}\right)(n-5)~.
\end{eqnarray}
In the $\mathbb Z_3$ orbifold analogous to that studied in the next section, the counting changes to
\begin{eqnarray}
E=
{3( -\tfrac1{12})+ 2(\tfrac1{36})}+
{8(\tfrac1{12})+ 8(-\tfrac1{36})}+
{n(-\tfrac1{12})+(11-n)(\tfrac1{36})}
=\left(-\tfrac 1{12}  -\tfrac1{36}\right)(n-5)~.
\end{eqnarray}
The $\mathbb Z_N$ orbifold works the same way with the factors $\frac1{36}$ replaced with the energy resulting from (\ref{ZPE}) with $\theta=\frac1N$. Since both orbifolds are different descriptions of the same smooth compactification, we see that $n=5$ is indeed the correct choice.
}
\begin{eqnarray}
E&=&\underbrace{6( -\tfrac1{24})+ 4(\tfrac1{48})}_{x+y}+
\underbrace{8(\tfrac1{12})+ 8(-\tfrac1{24})}_{\theta, p+ \theta^\prime ,p^\prime}+
\underbrace{n(-\tfrac1{12})+(11-n)(\tfrac1{24})}_{\lambda,w+ \lambda^\prime, w^\prime}\cr
&=&-\frac 18 (n-5)~.
\end{eqnarray}
The vacua which can contribute have energy $E=-\frac m2$ for $m=0, 1$, or $n=5, 9$ untwisted pure spinors, respectively.
However, as a negative energy right-moving vacuum would pair up with the negative energy left-moving one to make a tachyon, we are forced to consider only $n=5$.\footnote{This potential tachyon would be projected out if the vacuum were odd under the $\mathbb Z_2$ symmetry. This, however, would immediately lead to the wrong spectrum. Nevertheless, it is noteworthy that the additional ``solution'' exists. In the case of the CY 3-fold compactification, we will see even more such ``solutions''.} We are thus led to the conclusion that the ground state is of the form
\begin{eqnarray}
\Phi(x, \theta^{\alpha a}) \ket0_{vL}~,
\end{eqnarray}
where the wave function depends on the untwisted variables only.

We are missing some conditions which put this wave function on-shell. This is not surprising as we have only given a necessary condition for the existence of the twisted pure spinor vacuum. A more careful analysis requires the actual construction of the vacuum of the system resulting from solving the pure spinor conditions in terms of 11 unconstrained spinors and then twisting 6 of these. Such an analysis is beyond the scope of this work but we may use the existence and our knowledge of the vacuum to gain insight into the solution.

In order to couple the wave function above to the twisted states of the right-moving sector, we have to note that both states in (\ref{twistright}) are pseudoreal.  The second state of (\ref{twistright}) naturally couples to a doublet $\Phi_{a\bar{\jmath}}$ ($a$ is the doublet index)  satisfying the reality condition $(\Phi_{a\bar{\jmath}})^\ast= \epsilon_{ab}\epsilon_{\bar \imath \bar \jmath}\Phi^{b\bar \jmath}$. The first state in (\ref{twistright}) must couple to a doublet of pseudoreal superfields  $\Phi_{a x}$ which satisfies the reality condition $(\Phi_{a x})^\ast =\epsilon_{ab} C_{xy}  \Phi^{b y}$, where $C_{xy}$ is the charge conjugation matrix of the ${\mathbf {56}}$ representation. The correct representation is a half-hypermultiplet. This representation may be obtained from the cohomology of the operator (\ref{Q0}) in the ``reduced'' Hilbert space consisting of only the untwisted variables. This results in the condition that $\Phi$ be annihilated by the harmonic superspace derivatives $D^+_A$. Note that $\bm \lambda^\alpha$ is a pure spinor in six dimensions \cite{Berkovits:2005hy}. We will return to this in section \ref{IntMeas} where we will argue that the remaining untwisted component of the ten-dimensional pure spinor may be interpreted as a superspace harmonic.

In sum, each fixed point contributes a half-hypermultiplet transforming in the ${\mathbf {56}}$ of $E_7$  and a pair of half-hypermultiplets. Together with the states coming from the untwisted sector, we form the known spectrum. Note that two half-hypermultiplets can be combined to form the standard hypermultiplet.

\paragraph{Geometry and spectrum}
The Hodge theory surviving the orbifold projection is represented by the diagram
\begin{eqnarray}
\begin{array}{ccccc}
&&1&&\\
&0&&0&\\
1&&4&&1~~~.\\
&0&&0&\\
&&1&&
\end{array}
\end{eqnarray}
The six 2-forms $b_2= b_2^+ + b_2^-$ can be arranged into self-dual and anti-self-dual parts. The (anti-)holomorphic 2-form and K\"ahler form can be combined into a triplet of self-dual forms $\Omega_{ab}=\Omega_{ba}$. The remaining forms combine into a triplet of anti-self-dual forms $\Omega_{a^\prime b^\prime}=\Omega_{b^\prime a^\prime }$.

The orbifold points can be blown up to Eguchi-Hanson spaces $Y_v\cong \mathcal O_{\mathbb P^1}(-2) \cong T^*\mathbb CP^1$. The space $T^*\mathbb CP^1\sim S^2$ deformation retracts to the ``bolt'' so that the Hodge structure is given by
\begin{eqnarray}
\begin{array}{ccccc}
&&0&&\\
&0&&0&\\
0&&1&&0~~~.\\
&0&&0&\\
&&1&&
\end{array}
\end{eqnarray}
The single $(1,1)$-form $\Omega_v$ is anti-self-dual. The smooth K3 resulting from the blow-up will, therefore, have $b_2^+=3$ and $b_2^- = 3+16 = 19$.
The 16 $E_7$-neutral states $2\ket{(\mathbf 2,\mathbf 1,\mathbf 1)}$ from the twisted sector represent the orbifold limits of the $(1,1)$-forms on $Y$. Together with the $4$ neutral states surviving the projection in the untwisted sector, this gives $h^{1,1}=20$ for the K3 so that the Euler characteristic of the smooth K3 is $\chi=24$.

In the analysis above, we have embedded the spin connection in the gauge connection. Deforming away from this special limit can be done in 45 ways parameterized by elements of the cohomology $H^1(\mathrm {End}T)$ of holomorphic 1-forms with values in the endomorphism bundle of the tangent bundle.

\section{Four dimensions}
In this section, we repeat the six-dimensional analysis for the four-dimensional compactification of the pure spinor string. After reviewing four-dimensional results of \cite{Chandia:2009it}, we perform an orbifold operation and determine the massless spectrum. This time, matching to the known result in the twisted sector requires the twisting of $9$ pure spinors leaving $n=2$. Our conventions are summarized in appendix \ref{appendix4}.

\subsection{Supersymmetry in four dimensions}
In the compactification to four dimensions, the bosonic superspace coordinates split as $(X^m, Z^i, \bar Z^{\bar \imath})$ where $m=0, \dots, 3$ and $i,\bar \imath=1,2,3$ are (anti-)holomorphic indices. The fermionic superspace coordinates are $(\t^{\a}, \bar \t^{\dot \a},\t^{\a i}, \bar \t^{\dot \a \bar \imath})$ where $\a,\dot \a=1,2$ are $SL(2,\mathbb C)$ spinor indices. To avoid overly cumbersome notation, we identify the $SU(3)$ spinor indices with the $U(3)$ tangent space indices.
The pure spinor variables $(\l^{\a}, \l^{\dot \a},\l^{\a i}, \l^{\dot \a \bar \imath})$ are constrained by
\begin{eqnarray}
\label{PS1}
\lambda^\alpha \lambda^{\dot \alpha} + \lambda^{\alpha i}\lambda^{\dot \alpha}_i = 0
~,~
\lambda^{\alpha} \lambda_{\alpha}^i + \frac12 \epsilon^{ijk} \lambda^{\dot  \alpha}_{ j}\lambda_{\dot \alpha  k} =0
~,~
\lambda^{ \alpha i} \lambda_{ \alpha}^j + \epsilon^{ijk} \lambda^{\dot \alpha }\lambda_{\dot \alpha k} =0~.
\end{eqnarray}

As in ten dimensions, the pure spinor condition allows the definition of a nilpotent charge (\ref{qps}). Acting on massless states, this Berkovits differential determines the super-Maxwell equations of motion. In our case, the vector superfields $(A_m, A_i, \bar A_{\bar \imath})$ are related to the fermionic potential superfields $(A_{\a }, A_{\dot \a},A_{\a i}, A_{\dot \a \bar \imath})$ as
\be
D_\a A_\b + D_\b A_\a = D_\a A_{\bd i} + D_{\bd i} A_\a = D_{\ad} A_{\bd}+D_{\bd} A_{\ad}=0~,\cr\cr
D_\a A_{\b i}+D_{\b i}A_\a=-2i\varepsilon_{\a\b} A_i,\quad D_{\a i}A_{\b j}+D_{\b j}A_{\a i}=-2i\varepsilon_{\a\b}\e_{ijk}{\bar A}^k~,\cr\cr
D_\a A_{\bd}+D_{\bd}A_\a=-2i A_{\a\bd}, \quad D_{\a i}A_{\bd}^j+D_{\bd}^j A_{\a i}=-2i\d_i^j A_{\a\bd}~,\cr\cr
D_{\ad} A^i_{\bd}+D_{\bd}^i A_{\ad}=-2i\varepsilon_{\ad\bd}{\bar A}^i,\quad D^i_{\ad} A^j_{\bd}+D^j_{\bd}A^i_{\ad}=-2i\varepsilon_{\ad\bd}\e^{ijk}A_k~,
\ee
where
\begin{eqnarray}
\label{4D}
D_\alpha &=& {\partial \over \partial \theta^\alpha} + i \bar \theta^{\dot \alpha} {\partial \over \partial x^a}-i \theta_\alpha^i {\partial \over \partial y^i}~,\cr
\bar D_{\dot \alpha} &=& -{\partial\over\partial \bar \theta^{\dot \alpha}} -i \theta^\alpha  {\partial \over \partial x^a} +i \bar \theta_{\dot \alpha i} {\partial \over \partial \bar y_i}~,\cr
D_{\alpha i}&=& {\partial \over \partial \theta^{\alpha i}} + i \bar \theta^{\dot \alpha}_i {\partial \over \partial x^a} +i \theta_\alpha {\partial \over \partial y^i} - i \epsilon_{ijk}\theta^j_\alpha {\partial \over \partial \bar y_k}~,\cr
\bar D_{\dot \alpha}^i &=& -{\partial\over\partial \bar \theta^{\dot \alpha}_i} -i \theta^{\alpha i}  {\partial \over \partial x^a} -i \bar \theta_{\dot \alpha} {\partial \over \partial \bar y_i}+i \epsilon^{ijk} \bar \theta_{\dot \alpha j}{\partial \over \partial y^k}~.
\end{eqnarray}

As in six dimensions, we define covariant derivatives, which satisfy
\begin{eqnarray}
\label{CDAlgebra}
\hspace{-1.4cm}
\begin{array}{lll}
\{ \nabla_\alpha , \nabla_\beta\} = 0
	&\{\nabla_\alpha, \bar \nabla_{\dot \alpha}\}=-2i \nabla_a
	&\{\bar \nabla_{\dot \alpha} , \bar \nabla_{\dot \beta} \}=0\\
\{ \nabla_\alpha , \nabla_{\beta j} \} =- 2i \varepsilon_{\alpha \beta} \nabla_j
	&\{\nabla_\alpha, \bar \nabla_{\dot \alpha}^i\}=0
	&\{\bar \nabla_{\dot \alpha} , \bar \nabla_{\dot \beta}^j \}=-2i \varepsilon_{\dot \alpha \dot \beta} \bar \nabla^j~~~~~~.\\
\{ \nabla_{\alpha i} , \nabla_{\beta j}\} = -2i \varepsilon_{\alpha \beta} \epsilon_{ijk} \bar \nabla^k
	&\{\nabla_{\alpha i}, \bar \nabla_{\dot \alpha}^j\}=-2i \delta_i^j \nabla_a
	&\{\bar \nabla_{\dot \alpha}^i , \bar \nabla_{\dot \beta}^j \}=-2i \varepsilon_{\dot \alpha \dot \beta}\epsilon^{ijk}\nabla_k
\end{array}
\end{eqnarray}
At dimension-$\frac32$ we have the field strengths
\begin{eqnarray}
\begin{array}{lll}
\label{FSa}
[\nabla_\alpha , \nabla_b] = 2i \varepsilon_{\alpha \beta} \bar W_{\dot \beta}~,~
	&[\nabla_\alpha, \bar \nabla^i]=2iF_\alpha^i ,~,
&[\nabla_\alpha, \nabla_i]= 0~,
\end{array}
\end{eqnarray}
where we choose these normalizations for convenience. Due to the algebra, we can equivalently write them as
\begin{eqnarray}
\hspace{-1.0cm}
\begin{array}{lll}
\label{FSb}
[\nabla_{\alpha i} , \nabla_b] = 2i\varepsilon_{\alpha \beta} \bar F_{\dot \beta i} ,~,
		&[\nabla_{\alpha i}, \nabla_j]= -2i\epsilon_{ijk} F^k_\alpha ,~,
	&[\nabla_{\alpha i}, \bar \nabla^j]=-2i\delta_i^j W_\alpha~.
\end{array}
\end{eqnarray}
Proceeding to the dimension-2 field strengths, we define
\begin{eqnarray}
\begin{array}{lll}
[ \nabla_a , \nabla_b] = F_{ab} ~,~
	&[\nabla_a, \nabla_i]=F_{ai} ~,~
	&[\nabla_a , \bar \nabla^i ]=F_a{}^i~,\cr
[ \nabla_i , \nabla_j] = F_{ij} ~,~
	&[\nabla_i , \bar \nabla^j ]=F_i{}^j ~,~
	&[\bar \nabla^i,\bar \nabla^j ] = \bar F^{ij}~.
\end{array}
\end{eqnarray}
There are three additional {\it a priori} independent field strengths defined by
\begin{eqnarray}
d = \{ \nabla^\alpha, W_\alpha\}
	~,~ f^i = \{\nabla^\alpha , F_\alpha^i \}
	~,~\bar  f_i = \{\bar \nabla_{\dot \alpha} , \bar F^{\dot \alpha}_i\}~.
\end{eqnarray}
However, due to the two equivalent ways of writing the dimension-$\frac 32$ field strengths (\ref{FSa},\ref{FSb}),
these superfield strengths satisfy the Bianchi identities
\begin{eqnarray}
\label{Bianchis}
\hspace{-1cm}
\delta_i^j \{\nabla_{(\alpha} ,W_{\beta)}\}-\{\nabla_{(\alpha i}, F_{\beta)}{}^j\}=0~, &&
	\delta_i^j \{\nabla^\alpha, W_\alpha\}+ \{\nabla^{\alpha}_{ i},F_{\alpha}^{ j}\}-2 F_i{}^j = 0~,\cr
\{\nabla^{\alpha}_{ k} ,F_{\alpha}^{ k}\}+ F_k{}^k =0~,&&
	\{\nabla^{\alpha}_{ i}, F_{\alpha}^{ j}\}-2 F_i{}^j+ \delta_i^j F_k{}^k=0~,\cr
\{\nabla^\alpha, W_\alpha\} - F_k{}^k=0~,&&\\
\{\nabla_{(\alpha }, F_{\beta)}^k\} = 0~, &&
	\{\nabla^\alpha, F_\alpha{}^i\} - \epsilon^{ijk}F_{jk} = 0~,\cr
[\bar \nabla_{(\dot \alpha|}, F_{\alpha |\dot \beta)}{}^i]=0~,&&
	[\bar \nabla^{\dot \alpha}, F_a{}^i] +4i [\bar \nabla^i, W_\alpha]=0 ~,\cr
\{\bar \nabla_{\dot \alpha},F_\alpha{}^i\}+ F_a{}^i=0~,&&	
	[\nabla_i, W_\alpha ]+\frac i4 [\bar \nabla^{\dot \alpha},F_{ai}]-\frac12 [\nabla_a, \bar F^{\dot \alpha}_i ]= 0~,
\nonumber	
\end{eqnarray}
among which we find the equations
\begin{eqnarray}
\label{DF-HVB}
d= F_k{}^k~,~f_i=  \epsilon_{ijk}F^{jk}~,~\bar f^i=  \epsilon^{ijk}\bar F_{ij}~.
\end{eqnarray}
These are the $F$- and $D$-term conditions found in \cite{Chandia:2009it}. In the full superstring, they imply Ricci-flatness of the metric and the Hermitian-Yang-Mills equations on the connection.

\subsection{Calabi-Yau orbifold}
For purposes of illustration, we derive the result of orbifolding the 6-torus by the $\mathbb Z_3$ example presented in references \cite{Polchinski:1998rr} and \cite{Green:1987mn}. On the {\em right-moving} side we take
\begin{eqnarray}
\mathcal M(X_L^a,Z_L^i, \rho^{\mathrm i} , \rho^I, \rho^\mathbf I)
	= (X_L^a,\omega Z_L^i, \omega \rho^{\mathrm i} , \rho^I, \rho^\mathbf I)~,
\end{eqnarray}
where $\omega=\mathrm e^{\frac{2\pi i}3}$, $I=1,\dots, 10$, and $\mathbf I = 1, \dots 16$.
On the {\em left-moving} side,
\be
\mathcal M\mathcal (X_R^a,Z_R^i,\theta^\alpha , \theta^{\alpha i} , \lambda^\alpha , \lambda^{\alpha i})
	=(X_R^a, \omega Z_R^i,\theta^\alpha , \omega \theta^{\alpha i} , \lambda^\alpha , \omega \lambda^{\alpha i})~
\ee
together with their various conjugates.

\paragraph{Untwisted sector} The physics of the {\em right-moving} side is identical to the RNS analysis. The relevant states are built on
\begin{eqnarray}
\begin{array}{ccccl}
1 &:&\tilde \alpha_{-1}^a\ket0 &,& \ket{a}:a \in (\mathbf 8,\mathbf 1,\mathbf 1)\oplus(\mathbf 1,\mathbf {78},\mathbf 1)\oplus(\mathbf 1,\mathbf 1,\mathbf {248})~,\\
\omega &:&\tilde  \alpha_{-1}^i\ket0 &,& \ket{f}:f \in (\mathbf 3,\mathbf {27},\mathbf 1)~,\\
\omega^2 &:& \tilde \alpha_{-1}^{\bar \imath}\ket0 &,& \ket{\bar f} :\bar f \in (\overline{\mathbf 3},\overline{\mathbf {27}},\mathbf 1)~.
\end{array}
\end{eqnarray}
On the {\em left-moving}, side the 0-momentum cohomology generators are
\begin{eqnarray}
\begin{array}{cclclcl}
1&;&\Gamma^{\alpha \dot \alpha} \ket0
	 &,&\Gamma^{\alpha}\ket0 &,&\bar \Gamma^{\dot \alpha}\ket0~,  \\
\omega &;&\Gamma^{i}\ket0
	 &,&\Gamma^{\dot \alpha i}\ket0 &,& \\
\omega^2 &;&\bar \Gamma^{\bar \imath}\ket0
	 &,&\bar \Gamma^{\alpha \bar \imath}\ket0 &,&
\end{array}
\end{eqnarray}
where
\begin{eqnarray}
\begin{array}{ll}
\Gamma^{\alpha \dot \alpha} =i\left(
	 \bar \lambda^{\dot \alpha} \theta^\alpha - \lambda^\alpha \bar \theta^{\dot \alpha}
	 + \bar \lambda^{\dot \alpha}_{ i} \theta^{\alpha i} -\lambda^{\alpha i} \bar \theta^{\dot \alpha}_i\right)~,&
	\Gamma^{\alpha} = \bar \theta_{\dot \alpha} \Gamma^{\alpha \dot \alpha}+ \theta^{\alpha i} \Gamma_i~,\\
\Gamma^{i} =i\left(\lambda^{\alpha}\theta_{\alpha}^i - \lambda^{\alpha i}\theta_{\alpha}
		-\epsilon_{ijk} \lambda^{\alpha j} \theta_\alpha^k\right)~,&
	\Gamma^{\dot \alpha i}=\Gamma^i \bar \theta^{\dot \alpha} -\epsilon^{ijk} \bar \Gamma_j \bar \theta^{\dot \alpha}_k + \Gamma^a \theta_{\alpha}^i~,
\end{array}
\end{eqnarray}
and their conjugates.

The states surviving the projection form the supergravity multiplet and super-Yang-Mills multiplet in the adjoint representation of the unbroken gauge group $SU(3)\times E_6\times E_8$ from the $(1,1)$ combination. The $(\omega , \omega^2)$ combination gives $3\times 3 = 1+8$
neutral scalars and $3\times ({\mathbf 3}, \mathbf {27},\mathbf 1)$ scalars together with their superpartners. Finally, the $(\omega^2, \omega)$ sector gives the conjugates of these. We note in passing that the choice of chirality of the ten-dimensional pure spinor has resulted in the opposite of the usual convention for the choice of the ${\mathbf{27}}$-valued wave function, {\it id est}, chiral as opposed to anti-chiral.

\paragraph{Twisted sector} There are $3^3=27$ fixed points of the orbifold action. Each of these will carry a twisted-sector vacuum $\ket{0}_v= \ket{0}_{vL}\otimes \ket{0}_{vR}$.
Let us remind ourselves of the 0-point energies of the worldsheet fields as given by the formula (\ref{ZPE}). For the real periodic boson this is $-\frac1{24}$, for the real anti-periodic boson $\theta=\frac12$ it is $\frac1{48}$, for the real twisted boson with $\theta=\frac13,\frac23$ it is $\frac1{72}$, and for the real twisted boson with $\theta=\frac16$, $\frac56$, it is $-\frac1{144}$.

On the {\em right-moving} side, we again copy the RNS result.
The formula for the vacuum energy has a spacetime part which is always
\begin{eqnarray}
2(-\tfrac 1{24})+ 6(\tfrac1{72})=0~.
\end{eqnarray}
Here, we are taking into account the contribution of the $bc$ ghosts by counting only 2 of the 4 spacetime oscillators.
There are four sectors coming from the choice of periodic ($P$) and anti-periodic ($A$) boundary conditions for the heterotic fermions parameterizing the root lattice of $E_8\times E_8$. The fermions in the spin connection must have the same periodicity as the fermions in the commutant of $SU(3)\subset E_8$. Thus, there are, {\it a priori}, four sectors.
In the various sectors, we calculate the heterotic fermion contributions. The 6 fermions which were used to embed the spin connection into the gauge connection are twisted while the other $10+16$ heterotic fermions are not. This gives
\begin{eqnarray}
(P,P)&:& 6(-\tfrac 1{72}) +10(\tfrac1{24})+16(\tfrac 1{24})=1~,\cr
(A,P)&:& 6(\tfrac 1{144}) +10(-\tfrac1{48})+16(\tfrac 1{24})= \tfrac12~,\cr
(P,A)&:& 6(-\tfrac 1{72}) +10(\tfrac1{24})+16(-\tfrac 1{48})=0 ~,\cr
(A,A)&:& 6 (\tfrac1{144}) +10(-\tfrac1{48})+ 16(-\tfrac 1{48})=-\tfrac12~.
\end{eqnarray}

The vacua with positive mass do not contribute to the massless spectrum. The massless vacuum is acted on by $16-6=10$ fermion 0-modes $\rho_0^I$ which give the Pauli matrices of $Spin(10)$ upon quantization. As usual, this implies that the vacuum transforms in an irreducible spinor representation of $Spin(10)$, in this case the $\ket{PA}_{vR}\sim \overline{\mathbf {16}}$. Finally, the $(A,A)$ vacuum contributes the massless states
\begin{eqnarray}
\rho^{\mathrm i}_{-\frac16}\rho^{\mathrm j}_{-\frac16}\rho^{\mathrm k}_{-\frac16}\ket{AA}_{vR} \sim \mathbf{1}
~,~
\rho_{-\frac12}^I\ket{AA}_{vR}\sim \mathbf{10}
~,~
\rho_{-\frac16}^{\mathrm i}\tilde \alpha_{-\frac13}^{\bar \jmath} \ket{AA}_{vR}~.
\end{eqnarray}
The first two combine with the $\overline {\mathbf{16}}$ to give the $\overline{\mathbf {27}}$ of $E_6$. Together, these left-movers contribute
\begin{eqnarray}
\ket{(\mathbf 1, \overline{\mathbf {27}},\mathbf 1)}_{vR} + 3\ket{(\mathbf 3, \mathbf 1,\mathbf 1)}_{vR}~,
\end{eqnarray}
for every massless right-moving vacuum state, which we now construct.

On the {\em left-moving} side, we have $D=4$ untwisted and $10-D=6$ twisted coordinates. There are also $4N=4$ untwisted and $4(4-N)=12$ twisted fermions which come together with their conjugate momenta. Counting in pairs, these contribute $\frac1{12}$ and $-\frac1{36}$. Finally, we have $n$ untwisted and $11-n$ twisted pure spinors and their momenta giving $-\frac1{12}$ and $\frac1{36}$ in pairs. In all,
\begin{eqnarray}
E&=&\underbrace{4( -\tfrac1{24})+ 6(\tfrac1{72})}_{x+y}+
\underbrace{4(\tfrac1{12})+ 12(-\tfrac1{36})}_{\theta, p+ \theta^\prime ,p^\prime}+
\underbrace{n(-\tfrac1{12})+(11-n)(\tfrac1{36})}_{\lambda,w+ \lambda^\prime, w^\prime}\cr
&=&-\frac 19 (n-2)~.
\end{eqnarray}
The vacua which could, potentially, contribute have energy $E=-\frac m3$ for $m=0, 1,2, 3$, or $n=2,5, 8, 11$ untwisted pure spinors, respectively. However, as a negative energy right-moving vacuum would pair up the negative energy left-moving one to make a tachyon, we are forced to consider only $n=2$. We are thus led to the conclusion that the ground state is of the form
\begin{eqnarray}
\Phi(x, \theta, \bar \theta) \ket0_{vL}~,
\end{eqnarray}
where the wave function depends on the untwisted variables only.

As was the case in six dimensions, additional constraints are needed to impose the physical state conditions.
Again, this is not unexpected as we should really be computing the twisted vacua from a careful analysis of the pure spinor conformal field theory. As before, we will content ourselves with the by-hand imposition of the required conditions on the wave function.
The correct representation is that of a anti-chiral\footnote{As mentioned earlier, this unconventional result is due to the choice of chirality of the ten-dimensional pure spinor and the definition of the $\mathbf {27}$ ({\it v.s.}~the $\overline{\mathbf{27}}$). We have chosen to conform to the conventions of \cite{Berkovits:2000fe} for the former and those of \cite{Polchinski:1998rr} for the latter.} superfield $D_{\alpha}\bar \Phi= 0$, which corresponds to the calculation of the cohomology of the reduced differential
\begin{eqnarray}
Q_0 = \oint \lambda^{\alpha} d_{\alpha}~,
\end{eqnarray}
in the subspace without the twisted variables. Which $n=2$ pure spinor degrees of freedom we keep in the reduction of the Hilbert space should really be derived from a GSO condition. Here, we have simply picked the one that gives the correct answer that the field be anti-chiral (as opposed to chiral).\footnote{In the case of a type-IIA (resp.\,type-IIB) compactification, this solution implies that there will be one (twisted-)chiral field for each orbifold point $v=1,\dots, 27$. Since, as we will review below, these correspond to $(1,1)$-forms on a smooth CY resolution, we conclude that a smooth CY compactification will result in $h^{1,1}$ (twisted-)chiral fields.} The additional condition $\bar D^2 \bar \Phi=0$ is required to put the chiral field on-shell. This condition is also required in order to give the physical state conditions in a superfield description of all of the untwisted states found previously.

\paragraph{Geometry and spectrum} We have constructed a Calabi-Yau orbifold $\check X=T^6/\mathbb Z_3$ with 27 conical singularities. The Hodge diamond is
\begin{eqnarray}
\begin{array}{ccccccccc}
&&&1&&&\\
&&0&&0&\\
&0&&9&&0\\
1&&0&&0&&1~.\\
&0&&9&&0\\
&&0&&0&\\
&&&1&&&
\end{array}
\end{eqnarray}
Blowing $\check X$ up into a smooth CY 3-fold $X$ involves replacing the orbifold points with non-compact CY 3-folds $Y$. The simplest choice for $Y$ is $Y\cong\mathcal O_{\mathbb P^1}(-3)$ \cite{Green:1987mn}. Since $Y$ deformation-retracts to $\mathbb CP^2$, its cohomology is the same as that of the complex projective space:
\begin{eqnarray}
\begin{array}{ccccccccc}
&&&0&&&\\
&&0&&0&\\
&0&&1&&0\\
0&&0&&0&&0~.\\
&0&&1&&0\\
&&0&&0&\\
&&&1&&&
\end{array}
\end{eqnarray}
This gives $\chi(X)= 72$. The neutral states $3\ket{(\mathbf 3, \mathbf 1,\mathbf1)}$ become $(1,1)$-forms on $Y$ so there are $27$ of them in all. They parameterize the K\"ahler deformations on $X$.

In this analysis, the spin connection has been embedded in the gauge connection. The number of ways of doing this is parameterized by $H^1(\mathrm {End}TX)$. Changing the choice does not affect the topology of $X$ but it does affect the choice of complex structure. The number of complex structure deformations is, therefore, $\mathrm {dim} H^1(\mathrm {End} TX)$.

\section{Integration measures}
\label{IntMeas}
{Integration measures} are constructed by normalizing (a representative of) the unique element in the top level of the pure spinor cohomology (\ref{topform}). The construction requires regularization. In the non-covariant formalisms, this can be implemented by insertions of the picture-changing operator $Y= (C\cdot \theta)\delta(C\cdot \lambda)$. For the result of the $\lambda$-integration to be finite, we need ${c^\lambda}/2$ insertions where \cite{Berkovits:2005hy}
\begin{eqnarray}
c^\lambda= 2+\tfrac D2(\tfrac D2-1)~,
\end{eqnarray}
is the central charge of the pure spinor system in $D$ space-time dimensions.
Then, the measure is of the form
\begin{eqnarray}
\langle \lambda^{p} \theta^{q} Y^{\frac{c^\lambda}2}\rangle =1~.
\end{eqnarray}
It follows that the result of the $\theta$-integration is non-vanishing only if
\begin{eqnarray}
q=s-\frac{c^\lambda}2~,
\end{eqnarray}
where $s$ denotes the real dimension of the minimal spinor representation in $D$ dimensions. For $D=10$, $6$, and $4$, this is $s=2D-4$.
The $U(1)_J$ charge of this measure has to cancel the $U(1)_J$ anomaly $a$:
\begin{eqnarray}
p+q = -a~.
\end{eqnarray}
The ghost contribution to this anomaly is required to be
\begin{eqnarray}
a=D-2~,
\end{eqnarray}
in order to get the correct central charge \cite{Berkovits:2005hy}.
These considerations are summarized in the following table:
\begin{eqnarray}
\begin{array}{c|rrrrlrcc}
D	&{c^\lambda}/2&n	&s		&a		&c^\mathrm{matter}		 &p		&q	&\mathrm{measure}\\
\hline	
10	&11			&11			& 16		& -8		&-22								 &3 		&5		&\mathrm{pure~spinor}\\
6	&4 			&5			&8		&-4		&-10								 &0		&4		&\mathrm{analytic/chiral}\\
4	&2			&2			&4		&-2		&-4								 &0		&2		&\mathrm{chiral}
\end{array}
\end{eqnarray}

Note that the measures in $D=6$ and $4$ do not get pure spinor factors, as expected due to the existence of analytic and chiral measures in superspaces with $8$ and $4$ supercharges. Although there are no pure spinors in these measures, the 4, respectively 2, remaining pure spinors are required to define the analyticity/chirality constraints. Additionally, in the six-dimensional case, one more ghostlike degree of freedom is needed to cancel the central charge. It will appear as an unconstrained integration variable which suggests that it is a harmonic parameter. Since there is only one holomorphic such variable, it naturally appears as the projective parameter in the projective formulation of harmonic superspace \cite{Karlhede:1984vr}. This is consistent with the cohomological interpretation in section \ref{K3O} since there, the annihilation of a wave function by $Q_0$ implies that it is a projective superfield.

\section{Conclusions}

In the process of studying orbifold compactifications of the pure spinor superstring, we have encountered features which hint at an overall picture of general covariant compactifications. Previous work had shown that pure spinor conformal field theories can be formulated in six and four dimensions with 4 and 2 pure spinor degrees of freedom, respectively \cite{group,Berkovits:2005hy}. As they stand, such theories are not superstring theories as they fail to generate the physical state conditions.

We have shown that in the case of the compactification to six dimensions, the number of untwisted ten-dimensional pure spinors in the twisted sector has to be $n=5$ in order to match the spectrum and not generate tachyons. Furthermore, the twisted state wave functions must be constrained by a condition which coincides with that resulting from a differential constructed from a 4-component six-dimensional pure spinor. Finally, the 0-mode integration measure is expected to be independent of these 4 components but will include the integration over the $5-4=1$ additional bosonic degree of freedom. This further suggests that the additional variable can be interpreted as a projective \cite{Karlhede:1984vr} parameter on an orbit of the $SU(2)_L$ part of the internal Lorentz group. This is in addition to---and should not be confused with---the independent $u_{a^\prime}$ parameters which are harmonic-type parameters of the $SU(2)_R$ part which were introduced by hand in section \ref{K3O} to parameterize the choice of orthogonal complex structure.

The picture emerging in the four-dimensional case is simpler. 
The number of untwisted pure spinors in the twisted sector is $n=2$, in agreement with the expected number of four-dimensional pure spinors. The twisted state wave functions are chiral fields as one would obtain from a four-dimensional differential. Finally, the chiral integration measure emerges naturally from the regularized 0-mode normalization.

Despite the insights gained, our analysis leaves unanswered various important questions. Principal among these is the mechanism by which the physical state conditions are imposed directly on the cohomology of the lower-dimensional Berkovits differential. As we have approached the problem here, the conditions either come directly from the component spectrum (which admit no auxiliary component fields) or can be obtained by the method of reference \cite{Chandia:2009it}. Nevertheless, the emergence of a more covariant prescription is clearly preferable. Other topics meritorious of further study include the proper analysis of the pure spinor conformal field theory on the orbifold, the derivation of the pure spinor structure of the twisted vacua, the pure spinor vertex operators associated to the resolution of the orbifold singularities, and the covariant calculation of amplitudes in Calabi-Yau backgrounds.

\section{Acknowledgements}
We would like to thank Nathan Berkovits and Gabriele Tartaglino-Mazzucchelli for useful discussions. The work of WDL3 is supported by {\sc fondecyt} grant number 11100425 and {\sc dgid-unab} internal grant  DI-23-11/R. The work of BCV is partially supported by {\sc dgid-unab} internal grant DI-22-11/R.

\appendix
\section{Fierz identities in various dimensions}

In this appendix we collect some formulas involving gamma matrices in ten, six, and four dimensions.

\subsection{Ten dimensions}
\label{appendix10}
In ten dimension, Pauli-type matrices $(\gamma^m)_{\alpha \beta}$ are $16\times 16$ and symmetric matrices satisfying the Dirac algebra

\be
\g^m_{\a\g}(\g^n)^{\g\b}+\g^n_{\a\g}(\g^m)^{\g\b}=2\eta^{mn}\d_\a^\b~,
\label{tendirac}
\ee
and the Fierz identity
\begin{eqnarray}
\label{10dFierz1}
(\gamma^m)_{(\alpha \beta}(\gamma_m)_{\gamma) \delta} = 0~.
\end{eqnarray}

We define the completely antisymmetric products of gamma matrices as $\gamma^{m_1\dots m_p}=\frac 1{p!}(\gamma^{m_1}\cdots\g^{m_p}+\cdots)$ such that $\gamma^{m_1\dots m_p}=\gamma^{m_1}\cdots\g^{m_p}$ for $m_1\neq m_2\neq\cdots\neq m_p$. Any bi-spinor can be expanded in terms of a subset of these antisymmetric products. Note that $\g^m$ and $\g^{mnpqr}$ are symmetric while $\g^{mnp}$ is antisymmetric, then
\be
M_{\a\b}=\g^m_{\a\b}M_m+\g^{mnp}_{\a\b}M_{mnp}+
\g^{mnpqr}_{\a\b}M_{mnpqr}~,
\label{madbd}
\ee
where
\be
M_m=\frac{1}{16}\g_m^{\a\b}M_{\a\b} ~,~ M_{mnp}=\frac{1}{96}\g_{mnp}^{\a\b}M_{\a\b} ~,~ M_{mnpqr}=\frac{1}{2304}\g_{mnpqr}^{\a\b}M_{\a\b}~.
\ee
Similarly
\be
M_\a{}^\b=\d_\a{}^\b M + (\g^{mn})_\a{}^\b M_{mn}+(\g^{mnpq})_\a{}^\b M_{mnpq}~,
\label{madbu}
\ee
where
\be
M=\frac{1}{16}M_\a{}^\a ~,~ M_{mn}=-\frac{1}{32}(\g_{mn})_\a{}^\b M_\b{}^\a ~,~ M_{mnpq}=\frac{1}{384}(\g_{mnpq})_\a{}^\b M_\b{}^\a~.
\ee
With these expressions we can prove identities like
\begin{eqnarray}
\label{10dFierz2}
(\gamma^m)_{\alpha \delta} (\gamma_m)_{\gamma \beta}
       = -\frac12 (\gamma^m)_{\alpha \beta} (\gamma_m)_{\gamma \delta}
               +\frac1{24}(\gamma^{mnp})_{\alpha \beta} (\gamma_{mnp})_{\gamma \delta}~,
\end{eqnarray}
\begin{eqnarray}
(\gamma^{mn})_\alpha{}^\beta(\gamma_{mn})_\gamma{}^\delta = 4(\gamma^{m})_{\alpha \gamma} (\gamma_{m})^{\beta \delta}-2 \delta_\alpha^\beta \delta_\gamma^\delta -8 \delta_\alpha^\delta \delta_\gamma^\beta~.
\end{eqnarray}

\subsection{Six dimensions}
\label{6dsusy}

In six dimensions, the bosonic superspace coordinates split as $(X^m, X^i)$ where $m=0, \dots, 5$ and $i=6, \dots, 9$. The fermionic superspace coordinates are $(\t^{\a a}, \t^{a'}_\a)$ where $\a=1, \dots, 4$ are $SU(4)$ spinor indices and $a, a'=1,2$ are $SU(2)\times SU(2)$ spinor indices. It is not possible to raise or lower the $SU(4)$ spinor indices while the $SU(2)$ spinor indices can be raised or lowered with the antisymmetric symbols $(\e_{ab}, \e^{ab})$ and $(\e_{a'b'}, \e^{a'b'})$ in the following way
\be
\label{psia}
\psi^a=\e^{ab}\psi_b , \quad \psi_a=\e_{ab}\psi^b~,
\ee
similarly for primed indices. We use the convention $\e_{12}=-1, \e^{12}=+1$ such that $\e_{ab}\e^{bc}=\d_a^c, \e_{a'b'}\e^{b'c'}=\d_{a'}^{c'}$.

The non-zero components of the ten-dimensional gamma matrices of (1) in the six-dimensional language become
\be
\label{gm6d}
(\g^m)_{\a a,\b b}= \e_{ab}(\s^m)_{\a\b} ~,~ (\g^m)^\a_{a',}{}^\b_{b'}=\e_{a'b'}(\s^m)^{\a\b}~,
\ee
\be
\label{gi6d}
(\g^i)_{\a a},{}^\b_{b'}=(\s^i)_{ab'}\d_\a^\b~,
\ee
where $(\s^m)_{\a\b}$ and $(\s^m)^{\a\b}$ are the $SU(4)$ Pauli matrices which satisfy
\be
\label{pauli6d}
(\s^{(m})_{\a\g}(\s^{n)})^{\g\b}=2\eta^{mn}\d_\a^\b~,
\ee
and are related by $(\s_m)^{\a\b}=\frac{1}{2}\ve^{\a\b\g\d}(\s_m)_{\g\d}$. The matrices $(\s^i)_{aa'}$ and $(\st^i)^{a'a}$ are the $SU(2)\times SU(2)$ Pauli matrices which satisfy
\be
\label{pauli4d}
(\s^{(i})_{aa'}(\st^{j)})^{a'b}=2\eta^{ij}\d_a^b ~,~ (\s^{(i})_{aa'}(\st^{j)})^{b'a}=2\eta^{ij}\d_{a'}^{b'}~.
\ee
Similarly, the non-vanishing components of the ten-dimensional gamma matrices with upper indices are
\be
\label{Gm6d}
(\g^m)^{\a a,\b b}=\e^{ab}(\s^m)^{\a\b} ~,~ (\g^m)^{a'}_{\a,}{}^{b'}_{\b}=\e^{a'b'}(\s^m)_{\a\b}~,
\ee
\be
\label{Gi6d}
(\g^i)^{\a a}{},^{b'}_{\b}=(\st^i)^{b'a}\d_\a^\b~.
\ee
Note that the above sigma matrices satisfy the identities
\be
\label{ss6}
(\s^m)_{\a\b}(\s_m)_{\g\d}=\ve_{\a\b\g\d} ~,~ (\s^m)^{\a\b}(\s_m)^{\g\d}=\ve^{\a\b\g\d}~,
\ee
\be
\label{ss66}
(\s_m)^{\a\b}(\s^m)_{\g\d}=\d^\a_{[\g}\d^\b_{\d]} ~,~ (\s^i)_{aa'}(\s_i)_{bb'}=2\e_{ab}\e_{a'b'}~,
\ee
which come from the ten-dimensional Fierz identity.

\subsection{Four dimensions}
\label{appendix4}
In four dimensions, the Pauli gamma matrices are $\s^m_{\a\ad}$ and $\st_m^{\ad\b}$, where $m=0, \dots, 3; \a,\ad=1,2$. They satisfy,
\be
(\s^m\st^n+\s^n\st^m)_\a{}^\b=2\eta^{mn} \d_\a^\b ~,~ (\st^m\s^n+\st^n\s^m)_{\ad}{}^{\bd}=2\eta^{mn} \d_{\ad}^{\bd} ~.
\ee
Using this algebra, it is possible to show identities like
\be
\s_{\a\bd}\st^{\bd\a}= 2\eta^{mn} ~,~ \s^m_{\a\ad} (\s_m)_{\b\bd}=-2\epsilon_{\a\b}\epsilon_{\ad\bd}~,
\ee
where $\epsilon^{12}=-\epsilon_{12}=1$, similarly for $\epsilon$ with dottted indices. We use the identification $X_{\a\bd}=\s^m_{\a\bd} X_m$ between four-dimensional  bi-spinors and four-dimensional vectors.


\end{document}